\documentclass[11pt,a4paper]{scrartcl}

\usepackage{LCVision}

\usepackage{xpatch}
\makeatletter
\xpatchcmd\@HepConStyle
 {\edef\@upcode{\updefault}}
 {\ifdefined\shapedefault\edef\@upcode{\shapedefault}\else\edef\@upcode{\updefault}\fi}
 {}{}
\makeatother

\RedeclareSectionCommand[
  beforeskip=1.\baselineskip,
  afterskip=.5\baselineskip]{subsection}
\RedeclareSectionCommand[
  beforeskip=1.\baselineskip,
  afterskip=.5\baselineskip]{subsubsection}

\usepackage{todonotes}
\usepackage{xpatch}
\makeatletter
\xpretocmd{\todo}{\@bsphack}{}{}
\xapptocmd{\todo}{\@esphack}{}{}
\makeatother

\usepackage{booktabs}
\usepackage{csquotes}

\usepackage{graphicx} 
\usepackage{appendix}

\def\beq{\begin{equation}}
\def\eeq#1{\label{#1}\end{equation}}
\def\eeqn{\end{equation}}

\newenvironment{Eqnarray}%
   {\arraycolsep 0.14em\begin{eqnarray}}{\end{eqnarray}}
\def\beqa{\begin{Eqnarray}}
\def\eeqa#1{\label{#1}\end{Eqnarray}}
\def\eeqan{\end{Eqnarray}}

\let\bar=\overbar

\def\lsim{\mathrel{\raise.3ex\hbox{$<$\kern-.75em\lower1ex\hbox{$\sim$}}}}
\def\gsim{\mathrel{\raise.3ex\hbox{$>$\kern-.75em\lower1ex\hbox{$\sim$}}}}

\def\del{\partial}
\def\Dslash{\not{\hbox{\kern-4pt $D$}}}
\def\dslash{\not{\hbox{\kern-2pt $\del$}}}

\def\Dlr{\mathrel{\raise1.5ex\hbox{$\leftrightarrow$\kern-1em\lower1.5ex\hbox{$D$}}}}

\def\Pl{{\mbox{\scriptsize Pl}}}

\def\msb{{\bar{\scriptsize M \kern -1pt S}}}

\def\drb{{\bar{\scriptsize D \kern -1pt R}}}

\DeclareCiteCommand{\citejournal}[\mkbibbrackets]
  {\usebibmacro{prenote}}
  {\usebibmacro{citeindex}%
   \printtext[bibhyperref]{\printfield{journaltitle}}%
   \iffieldundef{volume}
     {}%
     {\setunit{\addspace}%
     \printtext[bibhyperref]{\printfield{volume}}}%
   \setunit{\addspace}%
   \printtext[bibhyperref]{(\printdate)}%
   \iffieldundef{pages}
     {}
     {\setunit{\addspace}%
     \printtext[bibhyperref]{\printfield{pages}}%
     }%
     }
  {\multicitedelim}
  {\usebibmacro{postnote}}
  
\DeclareCiteCommand{\citesubmit}[\mkbibbrackets]
  {\usebibmacro{prenote}}
  {\usebibmacro{citeindex}%
   \printtext[bibhyperref]{\printfield{journaltitle}}%
   \setunit{\addspace}%
   \printtext[bibhyperref]{(\printdate)}}
  {\multicitedelim}
  {\usebibmacro{postnote}}
  
  \DeclareCiteCommand{\citeconf}[\mkbibbrackets]
  {\usebibmacro{prenote}}
  {\usebibmacro{citeindex}%
   \printtext[bibhyperref]{\printfield{howpublished}}%
   \setunit{\addspace}%
   \printtext[bibhyperref]{(\printdate)}}
  {\multicitedelim}
  {\usebibmacro{postnote}}

\title{Contribution of\\ 
ALEGRO\\
to the Update of\\the European Strategy on Particle Physics 
}

\nolicence 

\notitlestamp 

\date{\today}

\addauthor{Editors: Brigitte Cros\footnote{brigitte.cros@cnrs.fr}}{\institute{1}}
\addauthor{Patric Muggli\footnote{muggli@mpp.mpg.de}}{\institute{2}}
\addauthor{on behalf of the ICFA-ANA panel members}

\addinstitute{1}{Laboratoire de Physique des Gaz et des Plasmas, CNRS, Universite Paris-Saclay, France}
\addinstitute{2}{Max Planck Institute for Physics, Garching/Munich, Germany}

\abstract{
\noindent
\textit{Abstract--}Advanced and novel accelerators (ANAs), driven a by laser pulse or a relativistic particle bunch, have made remarkable progress over the last decades. %
They accelerated electrons by 10\,GeV in 30\,cm (laser driven) and by  42\,GeV in 85\,cm (particle bunch driven). %
Rapid progress continues with lasers, plasma sources, computational methods, and more. %
In this document we highlight 
the main contributions made by the various major collaborations, facilities, and experiments that develop ANAs for applications to particle and high-energy physics. %
These include: ALiVE, ANL-AWA, AWAKE, BNL-ATF, CEPC Injector, DESY-KALDERA, ELI ERIC, EuPRAXIA, HALHF, LBNL-BELLA, LBNL-kBELLA, LCvison, PETRA IV Injector, 10\,TeV Collider design, SLAC-FACET II, as well as the development of structures, lasers and plasma sources, and sustainability, and demonstrate the intense activities in the field. %

ANAs can have, and already have, applications to particle and high-energy physics as subsystems, the so-called intermediate applications:
injectors, lower energy experiments, beam dump experiments, test beds for detectors, etc. %
Additionally, an ANA could be an upgrade for any Higgs factory based on a linear accelerator, as proposed in the LCvison project. %
ANAs have advantages over other concepts for reaching multi-TeV energies: lower geographical and environmental footprints, higher luminosity to power ratio, and are thus more sustainable than other accelerators. %

However, ANAs must still meet a number of challenges before they can produce bunches with parameters and the luminosity required for a linear collider at the energy frontier. %

It is therefore extremely important to strongly support vigorous R\&D of ANAs, because they are, at this time, \emph{the most sustainable acceleration scheme to reach very high energies with a linear accelerator}. %

They also have numerous lower energy applications as light and particle sources for research, industrial, medical, and security applications. %
    \begin{figure}[phtb]
    \centering
    \includegraphics[scale=0.5]{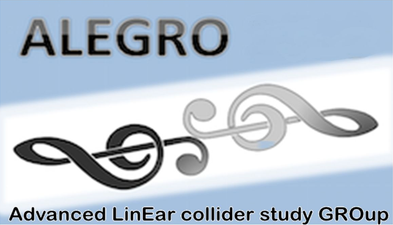}
    \end{figure}

}

\graphicspath{ {./logos/}{./figures/} }

\DeclareTOCStyleEntry[beforeskip=.2cm]{section}{section}

\begin{document}

\titlepage

\pagenumbering{arabic}\setcounter{page}{2}


\tableofcontents
\newpage
\section{List of Contributors}
\begin{itemize}
    \item Laura Corner, University of Liverpool, Liverpool, UK
    \item Brigitte Cros, Laboratoire de Physique des Gaz et des Plasmas, CNRS, Universite Paris-Saclay, France
    \item John Farmer, Max Planck Institute for Physics, Garching/Munich, Germany
     \item Massimo Ferarrio, INFN Frascati, Italy
    \item Spencer Gessner, SLAC National Accelerator Laboratory, Menlo Park, USA
    \item Leo Gizzi, Consiglio Nazionale delle Ricerche (CNR), Pisa, Italy
    \item Edda Gschwendtner, CERN, Geneva, Switzerland
    \item Mark Hogan, SLAC National Accelerator Laboratory, Menlo Park, USA
    \item Simon Hooker, Oxford University, Oxford, UK
    \item Wim Leemans, DESY, Hamburg, Germany
    \item Carl A. Lindstr{\o}m, University of Oslo, Oslo, Norway
    \item Jenny List, Deutsches Elektronen-Synchrotron, Hamburg, Germany 
    \item Andreas Maier DESY, Hamburg, Germany
    \item Patric Muggli, Max Planck Institute for Physics, Garching/Munich, Germany
    \item Jens Osterhoff, BELLA Center, Lawrence Berkeley National Laboratory, Berkeley, USA
    \item Philippe Piot, Argonne National Laboratory, Lemont, Illinois, USA
       \item John Power, Argonne National Laboratory, Lemont, Illinois, USA
    \item Igor Pogorelsky, Accelerator Test Facility, Brookhaven National Laboratory, Brookhaven, USA
    \item Marlene Turner, CERN, Geneva, Switzerland
    \item Jean-Luc Vay, Lawrence Berkeley National Laboratory, Berkeley, USA
    \item Jonathan Wood DESY, Hamburg, Germany
\end{itemize}
    ICFA-ANA Panel Members: Bruce Carlsten (LANL, USA), Brigitte Cros (CNRS, France), Massimo Ferrario (INFN, Italy), Simon Hooker (U. Oxford, UK), Tomonao Hosokai (U. Osaka, Japan), Masaki Kando (QST, Japan), Carl A. Lindstr{\o}m (U. Oslo, Norway), Patric Muggli (MPP, Germany, Chair), Jens Osterhoff (LBNL, USA), James Rosenzweig (UCLA, USA), Chuanxiang Tang (Tsinghua U., China).

 \newpage


\section{Introduction}

A 10\,TeV pCM energy is the grand challenge given to the accelerator community\footnote{Pathways to Innovation and Discovery in Particle Physics - P5 Report, Asai, S and others, 2024, \url{https://www.usparticlephysics.org/2023-p5-report/}.\label{P5}}. %
The collider must meet all requirements for HEP, but most of all, it must be sustainable and affordable. %
Operating at the largest possible accelerating gradient is therefore a great advantage for a linear lepton accelerator. %
Advanced and novel accelerators (ANAs) use structures or plasmas driven by laser pulses or particle bunches to reach large accelerating gradients. %
Gradients exceeding 30\,GeV/m to produce multi-GeV electron beams have been demonstrated in experiments (LBNL-BELLA, SLAC-FFTB, FACET and FACET II). %
The most recent experimental results show the production of a 10\,GeV electron beam in only 30\,cm of plasma. %

While ANAs have produced beams with some of the parameters required for a collider, producing these parameters together to reach the required luminosity remains a formidable challenge. %
However, no showstopper has been identified so far. %

ANAs are developed by a worldwide community that includes many independent research groups. %
This community includes university groups, national laboratories (ANL, BNL, CERN, DESY, LNF-INFN, LBNL, SLAC, ...), but also international collaborations (ALiVE\footnote{J. Farmer et al.,
New Journal of Physics 26 (2024) 113011.\label{ALiVE}}, %
AWAKE\footnote{\url{https://home.cern/science/accelerators/awake}.\label{AWAKE}},%
CAMPA\footnote{\url{https://campa.lbl.gov/}.\label{CAMPA}}, %
EuPRAXIA\footnote{\url{https://www.eupraxia-project.eu/}.\label{EuPRAXIA}}, %
HALHF\footnote{B. Foster et al., New Journal of Physics 25 (2023) 093037\label{HALHF}; E. Adli et al., HALHF: a hybrid, asymmetric, linear Higgs factory using plasma- and RF-based acceleration, 2025, \url{https://arxiv.org/abs/2503.19880}}, %
10TeV Collider Study\footnote{S. Gessner, J. Osterhoff et al., Design Initiative for a 10 TeV pCM Wakefield Collider, 2025, \url{https://arxiv.org/abs/2503.20214}.\label{10TeV}}, %
LCvison\footnote{J. List et al., A Linear Collider Facility for CERN, 2025, \url{https://indico.cern.ch/event/1471891/overview}.\label{LCvision}}, %
...). %
It is the mission of ALEGRO\footnote{ALEGRO Collaboration, Towards an Advanced Linear International Collider, 2019, \url{https://arxiv.org/abs/1901.10370}.\label{ALEGRO}}, %
the Advanced LinEar collider study GROup to: %
\begin{itemize}
    \item promote and encourage international collaboration/workshop/school on advanced and novel accelerators;
    \item especially emphasize advanced compact accelerators and their applications to not only high energy physics, particle physics, nuclear physics, but also medical physics, nondestructive evaluation, security etc.
\end{itemize}

While the ultimate goal of the community is an ANA-based collider at the multi-TeV energy level, \emph{in this document we highlight the diverse contributions ANAs can make as subsystems for, or intermediate applications to particle and high-energy physics, and to colliders.} %

Considering the promises of ANAs, \emph{it is essential to sustain a continuous support for their development, 
as recommended in the last ESPP update document:} %
"The technologies under consideration include ... plasma wakefield acceleration ...", among the ''innovative accelerator technologies'' that are part of the ''High-priority future initiatives''. %
However, we note that despite this positive inclusion and recommendation, no additional funding was made available, which greatly impedes progress towards the objectives highlighted by the plasma group. %

In this document, we briefly outline the relevant topics identified and addressed by the community. %
We give in the back-up document a list of the topics that major laboratories, projects or collaborations address in their own context. %

We note here that a clear physics case for Higgs factories has been developed in the context of ILC, CLIC, FCC-ee, CEPCee, HALHF, ALiVE, etc.

The physics case for a multi-TeV linear collider has been developed for CLIC, and is being continuously developed in the context of new projects such as LCvision %
and the 10TeV design effort. %

These are thus not discussed here, we focus on the accelerator itself. %

This contribution is based on the input of the contributors listed above, whose 19 inputs can be found in the back-up document. %

\section{Advanced and Novel Accelerators}
ANAs were proposed in the 80's\footnote{T. Tajima, et al., Phys. Rev. Lett. 43, 267 (1979), 	P. Chen, et al., Phys. Rev. Lett. 54, 693 (1985).} and are  still in their exploratory phase for applications to particle and high-energy physics. %

ANAs exploit the very large fields that dielectrics and plasmas can sustain. %
They use laser pulses and particle bunches to drive wakefields that are sustained only for a very short time, therefore increasing the amplitude they can reach without breakdown. %
Their operation at high gradient makes them extremely compact accelerators, mm- to cm-scale, to produce particles in the tens to hundreds of MeV energy range. %
These have multiple applications notably as particle sources for research and medicine, but also as radiation sources ranging from THz radiation to MeV photons. %

ANAs now produce beams with sufficient quality to drive free electron lasers (FELs). %
The next challenge is therefore that of a linear collider.

In the context of collider studies, an ANA has produced and accelerated electron bunches to 10\,GeV  in 30\,cm of plasma, i.e., at a gradient of 33\,GeV/m (LBNL-BELLA). %
They typically operate at high frequencies, and thus with smaller structures than other accelerators, requiring short, narrow driver and witness bunches. %
Highest frequencies and gradients are typically obtained in plasmas, where the accelerating structure is driven for every event. %
    \subsection{Grand Challenges}

    The amazing progress of ANAs in accelerating beams to multi-GeV energies encourages the community to address a number of remaining challenges in order to approach the requirements for a linear collider: 
    \begin{itemize}
        \item Emittance preservation and transverse stability
        \item Acceleration of positron bunches in plasma with collider parameters
        \item Staging of multiple accelerator plasmas
        \item Luminosity and efficiency
        \item Collider (including interaction point) with round beams favored by plasmas
    \end{itemize}
    These are seen as priorities and are all currently addressed by the community, either through global collider design efforts, or as single topics. %

    \subsection{Accelerator Components}
    An ANA-based collider has essentially the same main components as other colliders: injector, linac, beam-delivery system, and detector. %
    We briefly describe possible contributions of ANAs to these components. %
    ANA's have no contributions for detectors, but the ANA and the detector have to be developed together, as is done in ALiVE, HALHF, the 10TeV design, and LCvision projects. %

        \subsubsection{Injector}
        Plasma-based ANA operate with very small accelerating structure, requiring small and short witness bunches. %
        While RF-injectors can in principle be used (AWAKE), plasma-based injectors may provide witness bunches with parameters directly matching the needs of the accelerator: charge, emittance, duration, etc. %
        In particular, many schemes exist for trapping plasma electrons by and in the very large amplitude wakefields, thereby producing very low emittance bunches that may not require cooling in a damping ring. %
        A thorough study and optimization of plasma-based injector schemes was conducted in the context of EuPRAXIA and PETRA IV. %
        \subsubsection{Accelerating Components}
        
        ANA accelerating components are either plasma cavities created in the wake of the driver beam (particle or laser), or evacuated structures inside which wakefields are driven.
        
        \paragraph{Plasma Sources}
        ANAs operate with many different types of plasma sources, regarding gas confinement e.g., gas jet, gas cells, tubes, or plasma ionization methods e.g., discharge, Hydrodynamic Optical-Field-Ionization (HOFI) channel
        metallic vapor, etc. %
        Collider applications require controlled density distributions, reproducibility, operation at high repetition rate most likely implying cooling, etc. %
        Important efforts in developing these various sources take place for example in the context of AWAKE, LBNL-BELLA, EuPRAXIA, HALHF and at DESY-FLASHForward. %
        Laser-driven accelerators require optical guiding for the laser pulse for acceleration over long distances. %
        This is a major research topic and the continuous increase in energy gain was obtained mainly thanks to progress with optical guiding (LBNL-BELLA). %
        Improvements in plasma sources and laser beam quality (in particular phase front) are thus very actively pursued. %
        
        High-repetition rate, heating and cooling, and recovery of the plasma after an acceleration event were observed and studied experimentally at SLAC-FACET II and DESY-FLASHForward. %
        \paragraph{Structures}
        ANAs structures can be made of dielectric or metallic material. %
        Because they are driven by short particle bunches, they can sustain larger fields than structures powered by microwave sources. %
        In addition, dielectrics can sustain fields that approach the GV/m before breaking down. %
        Contrary to plasmas, they accelerate e$^+$ as well as e$^-$ bunches. %
        In general, while they sustain lower fields than plasmas, they could be the structures accelerating e$^+$ at high gradient in a collider that would accelerate e$^-$ in even higher gradient in plasma structures. %
        These structures are actively developed at ANL-AWA, and tested at a number of facilities (BNL-ATF, CERN). %
        We note that while HALHF uses RF- and plasma-based acceleration, a dielectric- and plasma-based collider could offer larger accelerating gradients and energies. %
        \subsubsection{Staging}
        To reach very high energies (e.g., $\gg$10\,GeV), plasma-based accelerators will require staging of multiple plasmas. %
        The large difference between the beta function of the beam inside and outside the plasma makes it necessary to  re-image the  accelerated bunch from the exit of the previous plasma onto the entrance of the next one. %
        Staging of two laser-driven plasmas is studied at LBNL-BELLA, while staging of multiple beam-driven plasmas is developed in the context of HALHF. %
        However, concepts that scale to the multi-TeV energies must be studied. %
        Staging is essential in maintaining the quality of the accelerated bunch (emittance preservation), as well as in maximizing the average accelerating gradient. %
        Indeed, the distance between stages is in general longer than the length of the accelerating plasma itself, and must be minimized while preserving beam quality. %
        
        \subsubsection{Laser Driver}
        Most lasers used today to power (plasma-based) ANAs are Titanium:Sapphire lasers. %
        Although these laser systems have been highly successful in demonstrations of ANAs to date, because of the large quantum defect (difference in pump and output laser photon energy), on the long-term these systems will not be able to meet the requirements for a collider driver: tens of kHz repetition rate, high average power (tens to hundreds of kW), mode quality, and, extremely importantly, high wall-plug efficiency. %
        Currently, to facilitate short- to medium-term accelerator research, Ti:Sapphire laser systems operating from 100\,Hz to a kHz are being developed at LNBL-kBELLA and DESY-KALDERA, and at a number of other institutions, as steps towards collider parameters. 
        For the longer term practical implementation of laser-driven plasma ANAs however, laser development research for plasma colliders is focused on laser systems beyond the current workhorse Ti:Sapphire. 
        A number of options are being actively investigated: 
        \begin{itemize}
        
    \item The generation of high intensity laser pulses by coherent combination of low energy but high efficiency, high repetition rate fiber lasers. 
    
    \item The investigation of other laser gain media with a smaller quantum defect than Ti:Sapphire. A leading option is thulium-doped yttrium lithium fluoride (Tm:YLF). %

    \item Development of alternative pumping for plasma ANAs. The use of multiple wavelength lasers creating a beat-wave in a plasma allows the use of longer pulse durations and thus alternative laser media.
    \end{itemize}
 %
        
        \subsubsection{Particle Beam Driver}
        An efficient particle beam driver system has been developed for CLIC. %
        A similar driver system could be suitable for the electron acceleration arm of the HALHF Higgs factory. %
        It could be used to power dielectric structures, whereas plasmas for very large energy applications may require higher beam energies to reach large energy gain per stage (10--20\,GeV). %
        \subsubsection{Beam Delivery System}
        The beam delivery system for a high-energy collider is many kilometers long. %
        Plasma lenses have been proposed to replace quadrupole magnets. %
        They provide focusing in the two planes simultaneously and have much larger focusing gradients than quadrupoles. %
        They may even be tapered to overcome the Oide limit. %
        These lenses are actively developed (SLAC-FACET II). %
        Beam delivery systems suitable for beam produced by ANAs are developed for the Higgs factory projects (ALiVE and HALHF) and in the 10TeV design project. %

        \subsubsection{Interaction Point}
        The classical physics at the interaction point can be described fully by the electromagnetic numerical particle-in-cell (PIC) simulation codes used to describe the acceleration process. %
        Many of these codes now include QED effects that have been added since one major application of plasma-based accelerators is strong-field QED, because of the  natural confluence of high-energy electron bunches and high-intensity laser pulses with ANAs. %
        There are thus many efforts to adapt these PIC "plasma" codes to include all processes occurring at the interaction point. %
        This is important because there may be indications that with the short, round and dense bunches produced by plasma-based accelerators, the luminosity to power ratio could be larger than with conventional flat beams, again increasing the sustainability of such a collider. %
        \subsubsection{Plasma Mirrors}
        A fundamental characteristics of plasma is that it reflects electromagnetic radiation with frequency lower than the plasma electron frequency. 
        A high-density  plasma, created at the surface of a material by the laser pulse that drives the plasma-based accelerator can be used very close to the entrance/exit of each stage to in-/out-couple the drive pulse, contributing to shortening the inter-stage section, i.e., maintaining a large average gradient. %
        These mirrors can be generated through ionization of nanometer-thick, very low vapor pressure liquid layers (liquid crystals) that minimize scattering of the accelerating bunch. %
        With a laser-driven accelerator the witness bunch travels straight through the in- and out-coupling mirrors, unlike in the beam-driven case, where magnetic chicanes wiggle the accelerating bunch at each inter-stage. %
        Such mirrors are developed at LBNL-BELLA and at  universities, in particular in the US and UK. %

        \subsubsection{Control System}
        Because of the strong interdependencies of the parameters in plasma-based accelerators, they can greatly benefit from automated optimization and control systems. %
        Therefore, many groups and laboratories (e.g., LBNL, DESY-FLASHForward, DESY-KALDERA, SLAC, SLAC-FACET II, etc.) develop and implement neural network (NN), machine learning (ML), and or Bayesian optimization systems. %
        These will play an essential role in reaching and maintaining parameters of the drive and accelerated beams. %

    \subsection{Intermediate Applications}
    Reaching collider parameters is the most stringent requirement for the accelerator. %
    However, there are a number of intermediate applications requiring more modest parameters that can be used as stepping stones for ANA-based acceleration schemes. %
    
        \subsubsection{Injector for a Circular Collider (or Light Source)}
        High-energy colliders usually use a large injector complex. %
        Particles are initially accelerated to energies in the range from the hundreds of MeV to the GeV energy level in linear accelerators before being injected in a circular accelerator. %

        Projects to build ANA-based injectors (in these cases, plasma-based) for a collider, CEPC
        , and for a light source (PETRA IV) 
        are under way. %
        \subsubsection{Fixed Target Experiments}    
        Many particle physics experiments use secondary beams produced for example by injectors for larger experiments, e.g., SPS beams at CERN. %
        However, the primary proton beam can be used to drive wakefields in plasma and accelerate a large number of electrons (or positrons) to energies comparable to that of the drive beam. %
        This energy gain can occur in a single plasma, avoiding again the intricacies of staging. %
        AWAKE is developing such an acceleration scheme and addresses a number of challenges common to many plasma-based ANAs: plasma source parameters, external injection, quality of the accelerated bunch, etc. %

        \subsubsection{Free Electron Lasers}    
        Though free electron lasers (FELs) are not directly collider instruments, except maybe in the context of a $\gamma\gamma$ collider, developing ANAs that produce beam parameters suitable to drive an FEL, in particular an X-ray FEL, is generally considered as an essential step towards collider applications. %
        Therefore, a number of university groups and laboratories actively work on ANAs for FELs: BNL-ATF, EuPRAXIA@SPARC\_LAB, LBNL, etc.

    \subsection{Collider Designs}
    A number of studies have taken on the challenge of the design of an ANA-based collider. %
    These include Higgs factories 
    and a 10\,TeV CM collider. %
    These are important because they include a physics case that then determines the acceleration scheme, e.g., ANA-based (10TeV, ALiVE) or mixed with RF acceleration (HALHF). %
    The 10TeV design has adopted a top-down approach that determines the parameters of one or combined ANAs optimizing particle physics requirements. %
    The choice of ANA was made for HALHF (beam-driven plasma for electrons) and ALiVE (proton-driven plasma) based on the availability of drivers. %
    A 10\,TeV ANA collider is considered as a possible upgrade for a lower energy linear collider in LCvison. %

    \subsection{The Positron Case}
    Acceleration of quality electron and positron beams is possible in structures. %
    However, plasma electrons sustain wakefields in plasmas, while ions are (assumed) immobile and retain a uniform density. %
    This leads to an asymmetry that allows for acceleration of quality electron beams, as was demonstrated many times, but not a priori that of positron beams. %
    While many schemes have been proposed to accelerate quality positron beams, none of them have been tested so far, mostly because suitable positron bunches are not available today. %
    
    Therefore, at this time, ANA-based accelerators focus on an e$^-$e$^-$ or $\gamma\gamma$ plasma-based collider, or on an e$^+$e$^-$ where electrons are accelerated in plasma, taking advantage of the largest accelerating gradient, while positrons are accelerated in a dielectric structure, taking advantage of their large gradient and e$^+$/e$^-$ symmetry. %

    Many possible schemes have been proposed and demonstrated in numerical simulations. %
    However, it is extremely important to make short, dense, relativistic positron bunches available (SLAC-FACET II) to test the proposed acceleration schemes and to stimulate new ideas. %

    \subsection{Numerical Simulations}
    Numerical simulations play a key role at all the stages of ANAs development. %
    Numerical simulation of plasma-based accelerators is particularly challenging with collider beam parameters. %
    This stems in particular from the nanometer transverse size of the bunch when matched to the focusing force of the wakefields. %
    This spatial scale is much smaller than that of the accelerating structure, requiring sophisticated mesh-refinement techniques. %
    Design and optimization can be performed with reduced models, in a boosted frame, and in 2D. %
    However, ultimately, fully-electromagnetic numerical simulations in 3D are necessary to confirm the results of the optimizations. %
    Many particle in cell (PIC) codes take advantage of GPU architectures to speed-up simulations. %
    A major global design effort is ongoing with the CAMPA project. %

    We note that every group and project involved in ANA development heavily uses numerical simulations to build and develop experiments, and to support the analysis and publication of experimental results. %
    In addition, most of the original ideas are first tested with numerical simulations. %
    
    \subsection{Global Concepts}
    Bringing ANAs from the status of experimental, single-event devices to that of accelerators for a collider requires the development of systems and concepts that meet much more stringent requirements. %

        \subsubsection{Efficiency and Luminosity}
        The efficiency of the driver system (laser or particle source), as well as that of the energy transfer from the driver and the witness are essential to minimize beam and site power, and cooling requirements for the accelerating structure. %
        The efficiency of these processes is also essential to maximize the sustainability of the collider. %
        Driver to witness bunch efficiency is addressed in experiments at DESY-FLASHForward and SLAC-FACET II. %
        These experiments are made possible by the availability of high energy electron bunches. %
        In the case of a laser diver, recycling of the energy remaining in the spent driver is essential. %
        This topic is addressed in many experiments, including at LBNL-BELLA. %

        Luminosity stems for the combination of single bunch parameters and high repetition rate, and maximizing it is thus intrinsically addressed in many experiments. %
        \subsubsection{Costing}
        Another essential criteria for the choice of a collider design is the availability of a fair and comprehensive costing model. %
        Such models are being developed for the ANA collider projects ALiVE and HALHF. %
        Costing is also used to prioritize development efforts. %
        \subsubsection{Sustainability}
        Maximizing sustainability is, of course, essential for all large size accelerators. %
        We have therefore included sustainability as one of the main topics at all level of ANA applications. %
        In fact, sustainability deriving from operation at high-gradient, i.e, smaller geographical footprint, may be the most attractive characteristics of an ANA-based collider. %
        However, all aspects (efficiency, impact of operation, building to dismantling approach, etc.) are now included in the various collider projects. %
        \subsubsection{Education and Training}
        Research and R\&D with ANAs occurs at many universities and national laboratories. %
        They mix many different sciences and technologies (plasmas, advanced materials, lasers, numerical simulations, etc.). %
        These, together with the novelty and potentials of ANA's attract many young scientists and engineers. %
        The field of ANA therefore acts as a large reservoir of creative and well-rounded engineers for universities, national laboratories, and industry to hire from. %
\section{Conclusions \& Outlook}
It is clear that the community developing ANAs for applications to particle and high-energy physics is very vibrant. %
It is also clear that rapid progress is being made. %
Whilst it is clear that many challenges remain to be met before the bunches and beams produced by ANAs meet the requirements for a collider, no showstoppers have been identified for the production of these bunches and beams. %
There are intermediate applications for which ANAs are competitive in terms of beam parameters and footprint. %

Most experiments with parameters relevant to particle and high-energy physics applications require multi-GeV beams, petawatt laser systems, and related large teams and equipment. %
These experiments are thus mostly performed at national-laboratory-scale-facilities, which means that funding and budgets must be commensurable. %
Contemplating the potential for these accelerators to contribute to particle and high energy physics and considering their advantages in the long term, \emph{it is essential to strongly support their further development}. %
\newpage\section{Contributions, Submitted to ESPP as "Back-up" for the Main Document}

This document features the input of the various contributors, input on which the main text is based. %
They are presented in alphabetical order of their title. %



\subsection{ALiVE, \textit{John Farmer}}
The ALiVE project seeks to extend the use of proton-driven wakefield acceleration to the high luminosities required for a future collider.  The use of a short proton driver allows witness acceleration to the energy frontier in a single plasma stage, avoiding the need to couple sequential acceleration sections.  The key areas of research offer significant synergies with other plasma acceleration schemes:
\begin{itemize}
  \item development of RF techniques for the acceleration of short proton bunches at high repetition rate
  \item optimization of beam and plasma parameters to control the quality of the accelerated bunches
  \item investigation of potential for high-quality positron acceleration in quasilinear wakefields
  \item development of plasma source technology for high repetition rates
  \item investigation and exploitation of high-disruption at the Interaction Point, arising from the acceleration of round beams in plasma
  \item scaling with drive beam energy, from Higgs factory to 10-TeV collider.
\end{itemize}

\subsection{Argonne AWA, \textit{Philippe Piot, John Power}}
The overarching mission of the Argonne Wakefield Accelerator (AWA) is to develop beam-driven acceleration techniques for an energy-frontier collider with a focus on structure-wakefield acceleration (SWFA) in both its collinear wakefield acceleration (CWA) and two-beam acceleration (TBA) implementations. Specifically, research at the 60-MeV AWA main beamline  includes:
\begin{itemize}
\item Generation and manipulation of drive beams including high-charge [${\cal O}(100~\text{nC})$] bunches  for TBA and development of advanced phase-space manipulations to improve efficiency in CWA techniques.
\item Design of accelerating structures capable of GV/m accelerating fields with operating frequencies  ranging from X-band (for TBA) to THz (for CWA).
\item Production of high-power Short RF pulses with a focus on power extracting and transfer structures capable of producing GW-scale peak powers with RF pulses $<10$~ns at X-band frequencies.
\item Integrated experiment to produce 500-MeV bright bunches using AWA as a drive-beam complex (in the TBA configuration)  with focus on beam-quality preservation, and ultimately on application to a free-electron laser operating in he water-window spectrum.
\end{itemize} 
To date, AWA has demonstrated reliable operation in a TBA configuration of both accelerating structures and photo-electron sources with surface fields approaching $\sim 0.5$~GV/m, currently limited only by available drive-beam bunch charge. A significant upcoming enhancement to AWA's capabilities is a planned energy upgrade to $\sim 120$~MeV, which will substantially expand experimental possibilities, enable exploration of higher-gradient acceleration regimes, and support the 500-MeV integrated TBA experiment.  

\subsection{AWAKE, \textit{Edda Gschwendtner, Patric Muggli}}
Main goal of AWAKE: develop a p$^+$-driven, plasma-wakefield accelerator to produce 50-100\,GeV e$^-$ bunches in a single plasma for application to particle physics (fixed target experiments), QED experiments, etc. %
The developments pursued that are shared with, or contribute (in general) to other plasma-based accelerators include:
\begin{itemize}
    \item External injection of an e$^-$ bunch $\Leftrightarrow$ injection, staging
    \item Plasma source: length, density uniformity, reproducibility, tunability $\Leftrightarrow$ staging, $\gg$1 plasma
    \item Acceleration to multi-GeV level $\Leftrightarrow$ large energy gain, HEP
    \item Control e-bunch quality: energy and energy spread $E$, $\Delta E$, charge $Q$, normalized emittance $\epsilon_N$ $\Leftrightarrow$ quality
    \item Diagnostics development: plasma density ($n_{e0}(z)$), bunch parameters ($\epsilon_N$) $\Leftrightarrow$ quality
    \item Control of instabilities: self-modulation, hosing, filamentation $\Leftrightarrow$ quality
    \item Control of motion of plasma ions $\Leftrightarrow$ quality
\end{itemize}

\subsection{Brookhaven ATF, \textit{Igor Pogorelsky}}
The Accelerator Test Facility (ATF) is the DOE Office of Science User Facility dedicated to the Accelerator Stewardship Program. %
It provides users with a comprehensive suite of capabilities, including a high-brightness electron LINAC synchronized with high-power Long-Wave IR (LWIR) and Near-IR lasers. %
These capabilities support a diverse range of studies related to Advanced and Novel Accelerators (ANA), such as:
\begin{itemize} 
    \item Two-color ionization injection for all-optical, high-brightness photoinjectors
    \item Plasma tomography for field mapping within laser-driven wake fields
    \item External injection of well-defined, low-emittance LINAC electron bunches with controlled energy spread into laser-driven plasma wakes
    \item Inverse Compton Scattering in novel regimes
    \item Investigation of "laboratory astrophysics," including plasma instabilities that seed magnetic fields in outer space
    \item Laser generation of monoenergetic ion beams from supersonic jets
    \item Beam-driven structure-based wakefield acceleration (SWFA) studies
    \item FELs and IFELs
and more ...
\end{itemize}
Leading the development of multi-terawatt, femtosecond LWIR laser technology, ATF is uniquely positioned to host detailed LWFA studies in a low plasma density regime where macroscopic-sized plasma bubbles, combined with precise e-beam and NIR laser injectors, might allow detailed investigations into LWFA seeding, staging, and bunch evolution during acceleration, pertinent to future colliders.


\subsection{CAMPA, \textit{Jean-Luc Vay}}
The Collaboration for Advanced  Modeling of Particle Accelerators (CAMPA) is a multidisciplinary, multi-institutional project funded by the U.S. Department of Energy, Office of Science, Office of Advanced Scientific Computing Research and Office of High Energy Physics, Scientific Discovery through Advanced Computing (SciDAC) program. The team includes accelerator physicists, applied mathematicians and software engineers from Lawrence Berkeley National Laboratory, Fermilab, UCLA, SLAC National Accelerator Laboratory, Argonne National Laboratory and Oak Ridge National Laboratory. 

The overarching purpose of CAMPA is to accelerate and expand the scope of discoveries from high energy physics (HEP) particle accelerators by enabling the design of accelerators that are significantly more compact and cheaper to build and run. 
To this end, the project has two main goals: (i) developing high-performance computing (HPC) accelerator and beam modeling capabilities to design the full range of systems required, from ``conventional'' to ``advanced concepts'' (e.g., plasma-based) and ``hybrids'' (mixes of conventional and advanced concepts), and (ii) developing community simulation ecosystems that seamlessly integrate accelerator elements to facilitate the design and control of next-generation accelerators.

One of the key targeted applications is the R\&D toward a plasma-based 10 TeV pCM energy collider design. CAMPA is working on developing further state-of-the-art simulation solutions, including PIC codes, workflows, standardized ecosystem, etc, that are needed for the study of such colliders. The tools are then used by CAMPA (or external collaborators with their own separate funds) to study a selection of topics of interest, including - but not limited to - beam generation and acceleration, stability, emittance preservation, staging, final focus system, beam crossing at the interaction point.


\subsection{DESY FLASHForward, \textit{Jonathan Wood}}
FLASHForward is an electron-bunch-driven plasma wakefield accelerator 
experiment at DESY. It capitalized on the FEL-grade electron beam and the high repetition rate of the superconducting RF accelerator to study beam-driven plasma acceleration. The main goals of FLASHForward are:

\begin{itemize}
\item The demonstration of high overall energy efficiency in a plasma wakefield 
accelerator
\item  Bunch-quality-preserving acceleration (charge, energy spread, emittance) 
with energy gains towards energy doubling of the accelerating bunch (from 
approx. 1 to 2 GeV)
\item  Acceleration of high-repetition-rate, high average power bunch trains 
(microsecond spacing), to demonstrate the compatibility of plasma boosters 
with superconducting RF linacs, as well as to inform general studies into 
high-repetition-rate plasma acceleration.
\item  To develop the diagnostics to measure electron bunch and plasma properties 
at these repetition rates
\item  To design and produce plasma targets capable of withstanding multi-kW 
(average) power deposition levels commensurate with the high average power 
operation of plasma accelerators.
\end{itemize}

In recent years, FLASHForward has progressed towards these goals by demonstrating:
\begin{itemize}
\item Per-mille level energy spread preservation with full charge coupling with 
50 MeV acceleration (2021)
\item 42\% instantaneous energy transfer efficiency, and separately 57\% driver 
energy depletion (2021 and 2024)
\item Emittance preservation and 3D brightness preservation of the accelerating 
bunch with an energy gain of 40 MeV (2024)
\item Demonstration that a plasma can recover to its initial state for an 
identical acceleration event to occur < 100 ns later (2022)
\end{itemize}

\subsection{DESY KALDERA, \textit{Andreas Maier, Wim Leemans}}
KALDERA is DESYs new laser to drive a new generation of laser-plasma accelerators at very high repetition rate. Based on the Ti:Sapphire architecture, KALDERA will eventually deliver 100TW class pulses at repetition rates of up to 1kHz. These high repetition rates will enable us to transfer well-known concepts of active stabilization and feedback from modern accelerators to the plasma world. Demonstrating sub-percent energy stability and, eventually, sub-percent energy spread GeV-scale electron beams over extended run times will help to build confidence in plasma technology in the accelerator community.
In a first project phase, KALDERA will run at 100Hz repetition rate. 20TW pulses are already available. First electron beams are expected for later 2025, which will be followed by 100TW peak power pulses at 100Hz later in 2026.
A new laser pulse compressor concept has been presented that can support kW class average power.


\subsection{EuPRAXIA, \textit{Massimo Ferarrio}}
EuPRAXIA (https://www.eupraxia-facility.org/) is an European project that develops a dedicated distributed particle accelerator research infrastructure based on novel plasma acceleration concepts driven by innovative laser and linac technologies. It is included in the ESFRI road map since 2021.  The EuPRAXIA@SPARCLAB facility is the beam driven pillar at LNF of the EuPRAXIA project. 
EuPRAXIA is expected to provide by the end of 2029 the first European Research Infrastructure dedicated to demonstrating usability of plasma accelerators delivering high brightness beams up to 1-5 GeV for users. 

The main goals of the project related to high energy , as stated in the EuPRAXIA Conceptual Design Report, are:
\begin{itemize}
    \item EuPRAXIA will deliver free-electron laser (FEL) X rays with $10^9-10^{13} $ photons per pulse to user areas, covering wavelengths of 0.2 nm to 36 nm. The EuPRAXIA FEL  will therefore provide users with tools for investigating processes and structures in ultra-fast photon science at a reduced facility foot print.



    \item  to deliver electron and positron beams at energies up to 5 GeV for high energy physics related R\&D (detectors, linear collider topics). R\&D goals include the demonstration of a linear collider stage, a "table top" HEP test beam and studies on positron transport and acceleration towards a linear collider.


      \item to develop a multi-stage, high-repetition rate plasma accelerator in the GeV range to users from accelerator science. This R\&D platform will allow the testing of novel ideas and concepts, full optimisation of a plasma collider stage, certain fixed target experiments (also in combination with lasers) and performance studies of conventional versus novel accelerator technology.

       \item EuPRAXIA will provide access to cutting edge high repetition rate laser and RF linac (X-band) technologies suitable for Synchrotron radiation and Linear Collider applications .
\end{itemize}
       Main current R\&D topics include:
\begin{itemize}
    
    \item Develop a concept and a proof of principle experiment for staging multiple plasma accelerator modules.

    \item Plasma accelerator theory and simulations.

    \item High repetition rate plasma module.

    \item High efficiency plasma acceleration, high transformer ratio test.
    
    \item Laser driven positron source and positron acceleration test.
\end{itemize}

\subsection{HALHF, \textit{Carl A. Lindstr{\o}m}}
The hybrid, asymmetric, linear Higgs factory (HALHF) is a plasma-based collider concept. The main goal of the project is:
\begin{itemize}
    \item A self-consistent conceptual design of an asymmetric plasma-based collider, based on electron-driven PWFA for high-energy electrons (e.g., 375 GeV) and RF acceleration for lower-energy positrons (e.g., 42 GeV), avoiding the ``positron problem" in plasma acceleration. Optimization of overall lifetime cost (construction + operation) guides the choice of technologies (for the driver production, positron acceleration and PWFA accelerators) as well as the choice of parameters. The increased concreteness allows new R\&D questions to be identified and tackled.
\end{itemize}
Main R\&D topics include:
\begin{itemize}
    
    \item Detailed concept for staging multiple plasma accelerator modules, including distribution of drivers.

    \item Repetition rate and heating limits of plasma accelerators and plasma cells.

    \item Preservation of spin-polarization in multistage plasma accelerators.

    \item Implementing required changes to ``conventional" collider subsystems: asymmetric detectors, higher-charge positron sources and damping rings, higher-emittance beam-delivery systems.
    
    \item Self-consistent, full-scale start-to-end simulations of a plasma-based collider, including modeling its cost (i.e., development of a ``system code").
    
\end{itemize}

\subsection{Laser Development, \textit{Leo Gizzi, Laura Corner}}
Here are a few example development items:

1) High quality, high average power, efficient pump lasers  using liquid cooling in place of gas cooling;

2) On the path to higher efficiency and higher repetition rate, thin disk picosecond Ytterbium (Yb) based systems with post-compression or other schemes (P-MOPA) or new Thulium (Tm) doped gain materials capable of broadband amplification and suitable for direct diode laser pumping. 

3) Among Tm doped materials, crystalline (YLF) or polycrystalline/ceramic (sesquioxides) media offer unique potential for efficient, scalable, very high repetition rate and average power. 

In view of these and other required developments, a shared strategic approach is being deployed with the PACRI Infratech EU project.
The PACRI project (Plasma Accelerator systems for Compact Research Infrastructures) was recently approved under the EU Horizon Work Programme 2023-2024, INFRA-2024-TECH-01-01. 

PACRI is a collaborative effort involving 26 partners (19 international research laboratories/universities and 7 EU industries) to develop highly relevant and breakthrough plasma accelerator technologies for Europe's future Research Infrastructures (RI). The official start of the project has been set for 1 March 2025.
PACRI includes 5 WP aimed at various aspects of development of efficient high repetition rate Laser drivers, including  High repetition rate high power Ti:Sapphire amplifier module, Efficient kHz laser driver modules for plasma acceleration, High-rep rate pump sources for laser drivers, Prototype of high average power optical compressor
and Laser driver system architecture, transport and engineering.
The programme sees major laser development laboratories with a leading role in these areas of development coming together with industrial partners.

\subsection{LBNL BELLA, \textit{Jens Osterhoff}}
Two main goals of BELLA for high-energy physics:
\begin{itemize}
    \item Development of a single-stage, proof-of-concept laser plasma accelerator at 10 GeV with high beam quality and high efficiency as a building block for a future 10 TeV collider
    \item Staging of two, beam-quality conserving, laser-plasma accelerator modules to achieve energies beyond 10 GeV
\end{itemize}
This research includes and is complemented by
\begin{itemize}
    \item Controlled injection and acceleration of high-brightness electron beams
    \item Development of laser guiding and plasma channels for high efficiency acceleration
    \item Development of novel, compact, electron beam transport concepts
    \item Development of concepts for the application of plasma accelerators to high-energy colliders
    \item Development of methods to model plasma accelerators
    \item Application of artificial intelligence / machine learning to laser-plasma acceleration for advanced control
    \item Transfer innovations in advanced accelerators and the emerging technology of plasma-based acceleration to applications in science, society and security.
\end{itemize}

\subsection{LBNL kBELLA, \textit{Jens Osterhoff}}
Main goal of kBELLA:
\begin{itemize}
    \item Development of ultrafast lasers and laser plasma accelerators at kHz+ repetition rate and kW class average power to advance laser and plasma technologies in a new facility towards the required high average power, efficiency, and precision control for pursuit of a compact 10 TeV parton center of mass (pCM) collider.
\end{itemize}
This includes:
\begin{itemize}
    \item The development of a kHz rep. rate, kW average power laser with the potential to scale the technology to >10 kW power in the future
    \item The realization of a test facility for high-average power operation
    \item The development of advanced control schemes at multi-kHz rates to enable precision, long-term stable laser plasma accelerators
    \item The application of these new capabilities to a multitude of applications in broad scientific fields, for society, and in the security space.
\end{itemize}

\subsection{LC Vision, \textit{Jenny List}}

The Linear Collider Vision initiative~\cite{LCVision-Generic} is taking a fresh look at the science goals and technologies for linear colliders, spanning from very mature, ready-to-build proposals, like ILC~\cite{ILC-EPPSU:2025, ILCInternationalDevelopmentTeam:2022izu} or CLIC~\cite{Adli:ESU25RDR, Brunner:2022usy}, via advanced SCRF and cool copper cavities to wakefield acceleration and energy and particle recovery schemes. It proposes in particular a linear collider facility (LCF) for CERN~\cite{LCF:EPPSU} which could perform a comprehensive program to study the Higgs boson and the top quark based on SCRF technology. At the same time, the LCF should offer R\&D facilities for novel accelerator technologies, in particular PWA, in order to prepare an upgrade of the facility with what turns out to be the most suitable one over the next 10-15 years. Options studied in the LCVision team reach from a HALHF-like upgrade to double the centre-of-mass energy for the case PWA of positron should not advance quickly enough for a  symmetric upgrade, to a 10\,TeV full PWA collider either for $e^+e^-$ or $\gamma\gamma$ collisions. The synergies of such an approach and example timelines are discussed also in~\cite{flexible-EPPSU:2025}. The concept of a linear collider facility -- at CERN or elsewhere -- thus offers an attractive and flexible vision for the future of particle physics for pursuing a precision study of the Higgs boson in $e^+e^-$ collisions at the earliest possible date while allowing to maintain or even intensify our effort in accelerator R\&D towards the 10\,TeV scale.

\subsection{PETRA IV Injector, \textit{Andreas Maier, Wim Leemans}}
DESY is currently exploring the option of a laser-plasma-based injector for its future PETRA IV synchrotron light source. A conceptual design report has recently been published. The concept is based on a PW-class drive laser that, combined with hydrodynamic optically-field-ionized (HOFI) guiding channels, provides 6GeV electron beams from a single plasma stage. To enhance the spectral properties, the laser-plasma electron beam will be actively energy compressed using a chicane in combination with a small correcting RF cavity. Currently, a prototype laser-plasma-based injector for DESYs current booster synchrotron (DESYII) is being built supported by a dedicated grant from German BMBF. The goal is to demonstrate injection into the booster ring by end of 2027. We see the successful deployment of plasma technology for light sources as an important stepping stone towards future high-energy physics applications. The strict requirements on availability and stability set forward by a photon science user community will motivate the same technological developments that are ultimately also necessary for operating a plasma-based collider.

\subsection{Plasma Sources Development, \textit{Simon Hooker}}
A wide range of work is being undertaken to develop plasma sources for both laser- and beam-driven plasma accelerators. This includes:

\begin{itemize}
        
    \item Development of novel gas cells or gas jets capable of kilohertz operation, which present a reduced gas load to the vacuum system.
    
    \item Development of all-optical laser guiding channels suitable as multi-GeV plasma stages and capable of high pulse repetition rates ($>$ 1\,kHz).
    
    \item Development of capillary discharge plasma channels for laser guiding.
    
    \item Development of solid-state discharge technology capable of MHz burst mode for driving capillary discharge plasma sources for beam-driven plasma accelerators.

    \item Development of 10-m-scale novel RF-heated and pulsed discharges for proton-driven plasma accelerators.
\end{itemize}

\subsection{SLAC FACET-II, \textit{Mark Hogan}}
FACET-II is a National User Facility at the SLAC National Accelerator Laboratory that has a broad User program based on high-energy high-peak-current electron beams and their interaction with lasers, plasmas and solids. Roughly fifty percent of the User program consists of collaboration-driven experiments designed to address milestones for beam-driven plasma wakefield acceleration (PWFA) outlined in the 2016 DOE Advanced Accelerator Development Strategy Report. \\

For PWFA, R\&D focuses on:
\begin{itemize}
    \item Demonstrating a single PWFA stage with 10\,GeV of energy gain with more than 100\,pC of accelerated charge and simultaneously having high-efficiency, narrow energy spread and preserved emittance
    \item Understanding the mechanisms that lead to emittance growth in the plasma such as mis-matching, chromaticity and hosing while demonstrating mitigation mechanisms such as plasma density ramps and ion motion
    \item Developing meter scale plasma sources that provide the appropriate transverse size and uniformity with tailored longitudinal density ramps
    \item Developing passive plasma lenses for matching into the plasma, as tools to reduce the interstage distance, and as potential stand-alone devices that might find application in collider beam delivery systems
    \item Optically probing the plasma to characterize the long-term density evolution and energy flow to identify optimal configurations for high repetition rate operation suitable for collider applications
    \item Developing ultra-high brightness electron sources based on injecting electrons directly into the high-gradient strong-focusing environment within the non-linear plasma wakefield. Injection mechanisms being studied include density downramp, plasma photocathode and ionization injection
\end{itemize}

\subsection{Structure Wakefield Acceleration, \textit{Philippe Piot}}
Structure Wakefield Acceleration (SWFA) offers two distinct implementation pathways: a semi-conventional approach based on two-beam acceleration (TBA), similar to the CLIC design architecture, and a beam-driven collinear-wakefield acceleration (CWA) method that parallels plasma-wakefield acceleration (PWFA) techniques. While the accelerating structures required for both implementations share fundamental design principles, they differ significantly in their operating frequencies, with TBA typically operating at lower frequencies (X and K bands) and CWA pushing into the THz regime comparable to PWFA.\\

For TBA, R\&D on structures focuses on:
\begin{itemize}
    \item Developing high shunt impedance power-extracting and transfer structures capable of generating GW-scale peak power with short duration and precisely tailored RF pulse envelopes
    \item Engineering high-gradient accelerating structures capable of sustaining GV/m peak fields with minimal breakdown rates and optimal efficiency
    \item Understanding dynamical effects in short-pulse acceleration, including potential impacts and mitigation strategies for detrimental effects on main-bunch beam dynamics and brightness preservation
    \item Designing integrated acceleration modules capable of GeV-class energy gain with compact footprints, including coupling mechanisms, alignment systems, and drive-beam complex
\end{itemize}

For CWA, R\&D on structures includes: 
\begin{itemize}
    \item Developing advanced beam shaping techniques to optimize the drive beam distribution for maximum wakefield excitation while minimizing deleterious effects
    \item Engineering precision timing and synchronization systems for drive-witness beam pairs with femtosecond-level stability to ensure optimal energy transfer
    \item Creating optimal bunch-train/current profile configurations that enable transformer ratio enhancement beyond the traditional limit of 2, potentially reaching values of $\sim 10$ for improved efficiency
    \item Investigating beam-loading compensation methods to preserve witness beam quality and minimize energy spread in high-gradient acceleration regimes
    \item Designing compact, integrated system with strong external focusing to explore onset and demonstrate mitigation of the single-bunch beam-break-up instability.
\end{itemize}

\subsection{Sustainability, \textit{Marlene Turner}}
Main Goal:\\
Minimize the environmental footprint of future accelerators and colliders based on advanced acceleration concepts.  
\\
\\
Approach:
\begin{itemize}
    \item Assess environmental impact and material use during construction and decommissioning following CERN LDG guidelines.  
    \subitem Identify harmful substances and explore alternatives.
    \subitem Minimize facility size. 
    \item Optimize power requirements during operation
    \subitem Investigate energy-saving measures such as energy recovery systems and beam parameter optimization.
    \subitem Evaluate different power sources and their environmental impact.
    \item Embed environmental considerations into the design process from the beginning.
    
\end{itemize}

\subsection{The 10TeV Wakefield Collider Design Study, \textit{Spencer Gessner}}
We have initiated the 10 TeV Wakefield Collider Design Study as a global endeavor in response to the P5 request~\cite{P5:2023}. 
The Design Study includes Working Groups for the components of the collider, which covers Particle Sources, Laser and Beam Drivers, Wakefield Linacs, the Beam Delivery System, and Beam-Beam Interactions. The Working Groups look to optimize the individual subsystems within a framework that imposes self-consistent beam parameters and concurrence between components of the accelerator. The Systems and Optimization Working Group will evaluate collider design choices with the goal of minimizing cost and environmental impact while maximizing the physics reach of the collider.


The Design Study exists within a larger framework of initiatives in pursuit of future colliders. The 10 TeV Wakefield Collider is an energy frontier upgrade path for a Linear Collider facility, as detailed in LC Visions~\cite{LCVisions:2025}. The 10 TeV Design Study operates in close collaboration with ALEGRO~\cite{ALEGRO:2019} and with the HALHF~\cite{HALHF:2023}, ALIVE~\cite{ALIVE:2024}, and XCC~\cite{XCC:2023} design studies. 
The 10 TeV Wakefield Collider Design Study leverages the ongoing efforts of these other studies (including the Muon Collider~\cite{MC:2024}) in order to make rapid progress and avoid redundant effort.

The deliverables for the 10 TeV Wakefield Collider Design Study include:
\begin{itemize}
    \item A description of the Physics Program for 10 TeV $e^+e^-$, $e^-e^-$, and/or $\gamma\gamma$ collisions.
    \item An end-to-end design concept, including cost scales, with self-consistent parameters.
    \item Roadmaps and resource estimates for medium-term R\&D.
\end{itemize}
The Design Study, in conjunction with ongoing experimental progress, will provide critical input for the High Energy Physics community as it evaluates viable and affordable paths toward 10\,TeV pCM collisions.

\newpage 


\end{document}